\documentclass[12pt,preprint]{aastex}
\usepackage{emulateapj5}



\newcommand{\lsim}{\mbox{\hspace{.2em}\raisebox{.5ex}{$<$}\hspace{-.8em}\raisebox{-.5ex}{$\sim$}\hspace{.2em}}}

\def\asca       {{\em ASCA}\/}
\def\chandra    {{\em Chandra}\/}

\def\xmm        {XMM-{\em Newton}\/}
\def\rosat      {{\em ROSAT}\/}
\def\sax        {{\em BeppoSAX}\/}

\def\mydegree{$^\circ\mskip-5mu$}
\def\myarcmin{$^\prime\mskip-5mu$ }
\def\myarcsec{$^{\prime\prime}\mskip-5mu$}
\def\rv         {$r_{\rm vir}$}

\usepackage{mathptmx}

\begin{document}

\title{\chandra\ observations of the NGC~1550 galaxy group ---
implication for the temperature and entropy profiles of 1 keV galaxy groups
}

\author{
M.\ Sun,$^{\!}$\altaffilmark{1}
W.\ Forman,$^{\!}$\altaffilmark{1}
A.\ Vikhlinin,$^{\!}$\altaffilmark{1}
A.\ Hornstrup,$^{\!}$\altaffilmark{2}
C.\ Jones,$^{\!}$\altaffilmark{1}
S. S. Murray$^{\!}$\altaffilmark{1}
} 
\smallskip

\affil{\scriptsize 1) Harvard-Smithsonian Center for Astrophysics,
60 Garden St., Cambridge, MA 02138; msun@cfa.harvard.edu}
\affil{\scriptsize 2) Danish Space Research Institute,
Juliane Maries Vej 30, 2100 Copenhagen O, Denmark.} 

\begin{abstract}

We present a detailed \chandra\ study of the galaxy group
NGC~1550. For its temperature (1.37$\pm$0.01 keV) and velocity
dispersion ($\sim$ 300 km s$^{-1}$), the NGC~1550 group is one of the
most luminous known galaxy groups (L$_{\rm bol}$ =
1.65$\times$10$^{43}$ erg s$^{-1}$ within 200 kpc, or 0.2 \rv). We find
that within $\sim 60$ kpc, where the gas cooling time is less than a
Hubble time, the gas temperature decreases continuously toward the
center, implying the existence of a cooling core. The temperature
also declines beyond $\sim$ 100 kpc (or 0.1 \rv). There is a remarkable
similarity of the temperature profile of NGC~1550 with those of two
other 1 keV groups with accurate temperature determination. The
temperature begins to decline at 0.07 - 0.1 \rv, while in hot clusters
the decline begins at or beyond 0.2 \rv. Thus, there are at least 
some 1 keV groups that have significantly different temperature profiles
from those of hot clusters, which may reflect the role of non-gravitational
processes in ICM/IGM evolution. NGC~1550 has no isentropic core in its entropy
profile, in contrast to the predictions of `entropy-floor' simulations.
We compare the scaled entropy profiles of three 1 keV groups (including
NGC~1550) and three 2 - 3 keV groups. The scaled entropy profiles of
1 keV groups show much larger scatter than those of hotter systems,
which implies varied pre-heating levels.
We also discuss the mass content of the NGC~1550 group and the
abundance profile of heavy elements.

\end{abstract}

\keywords{galaxies: groups: individual (NGC~1550) --- hydrodynamics ---
   X-rays: galaxies: clusters}

\section{Introduction}

In the simplest structure formation model, the thermodynamic
properties of the intracluster medium (ICM) are simply governed
by shock heating in gravitational infall. Thus, self-similar
relations, between mass, temperature and luminosity, can be expected
for different halos. However, X-ray observations show the observed
L$_{\rm x}$ - T relation of clusters, which is approximately L$_{\rm x}$
$\propto$ T$^{2.6}$ (e.g., Markevitch 1998), is not consistent with the
predictions of the self-similar model (L$_{\rm x}$ $\propto$ T$^{2}$).
The L$_{\rm x}$ - T relation for galaxy groups may be even steeper than
that of clusters (e.g., Helsdon \& Ponman 2000, HP00 hereafter). Ponman,
Cannon \& Navarro (1999) and Lloyd-Davies, Ponman \& Cannon (2000) found
that the central entropy of galaxy groups is higher than predicted by
self-similar scaling. These deviations from the self-similar model imply
departures from scale-free relations in actual ICM evolution.

The general consensus is that some non-gravitational processes must be
included in ICM evolution. Three ideas are currently
discussed in the literature: pre-heating (or star-formation feedback),
radiative cooling, and galaxy formation efficiency. For detailed
descriptions of the current understanding of these processes in
galaxy clusters and groups, see Tozzi \& Norman (2001, TN01 hereafter),
Babul et al. (2002) and Ponman et al. (2003).
Galaxy groups are ideal targets to study non-gravitational
effects since the implied non-gravitational energy is comparable to
the gravitational energy in these cool systems. However, pre-\chandra\
measurements generally have large uncertainties (e.g., Ponman et
al. 1999; Lloyd-Davies et al. 2000; Finoguenov et al. 2002).
The launch of \chandra\ and \xmm\ provide much stronger tools
to achieve precise measurement of temperature, abundance and
density profiles with which mass and entropy
profiles can be derived and L$_{\rm x}$ - T, M - T and S - T relations
can be better constrained. These well-established
relations, in combination with simulations, can reveal the roles
of non-gravitational processes in ICM evolution.
Among observational properties of galaxy groups, the temperature profile
is the most important but is still poorly constrained. It not only is
crucial to derive the mass and entropy profiles, but also reflects the
roles of non-gravitational processes and serves as an important
test of simulations (e.g., Muanwong et al. 2002; Dav$\acute{\rm e}$
et al. 2002).

We present in this {\it Paper} a detailed analysis of the NGC~1550
group, reported as a group by Garcia (1993). NGC~1550 was selected as a candidate
``Over Luminous Elliptical Galaxy'' (OLEG, Vikhlinin et al. 1999)
from a sample of early-type galaxies based on the \rosat\ All-Sky Survey
(Beuing et al. 1999). NGC~1550 (or NGC~1551, UGC 3012) is a lenticular galaxy at
z=0.0124. It was detected as a bright X-ray source in the
\rosat\ All-Sky Survey. The follow-up \rosat\ HRI observation revealed a
very extended halo ($>$ 15$'$ in radius), implying the existence of a
galaxy group/cluster. However, NGC~1550 dominates the optical light. There
was no previous information on the temperature of the system. NGC~1550
is a weak radio source (16.6$\pm$1.6 mJy at 20 cm) with a lobe-like
extension to the west revealed by the NRAO VLA Sky Survey (NVSS).
Thus, the nuclear radio output may not be important for the group
thermodynamics at the current stage. We will only discuss the global
properties of the NGC~1550 group in this paper, with emphasis on
the temperature profile and the comparison with other groups/clusters
and simulations. The central region will be discussed in detail
in a subsequent paper.

The plan of this paper is as follows: in $\S$2 we describe the \chandra\
observations and data analysis, including the global and radial properties
of the NGC~1550 group. In $\S$3 we describe its optical properties.
$\S$4 discusses the mass profiles. $\S$5 is the discussion and $\S$6 is the
conclusion. Throughout this paper we assume H$_{0}$ = 70 km s$^{-1}$ Mpc$^{-1}$,
$\Omega$$_{\rm M}$=0.3, and $\Omega_{\rm \Lambda}$=0.7. At a redshift
z=0.01239, the luminosity distance to NGC~1550 is d$_{\rm L}$ = 53.7 Mpc, and
1$''$=0.254 kpc. All physical scales, luminosities, densities, entropies,
gas masses, stellar masses, and total masses scale as d, d$^{2}$,
d$^{-1/2}$, d$^{1/3}$, d$^{5/2}$, d$^{2}$, and d, respectively, where
d=d$_{\rm L}$/53.7 Mpc.

\section{\chandra\ data analysis}

NGC~1550 was observed on January 8, 2002 by \chandra\ with the Advanced
CCD Imaging Spectrometer (ACIS). The observations were divided into two
ACIS-I pointings (Obs 3186 and 3187), with exposure times of 10.1 ks
and 9.8 ks, respectively. NGC~1550 was placed near the center of the I3
and I1 chips (Obs 3186 and 3187). In each
pointing, the aimpoint was moved along the detector Y axis to be
closer to the target. This provides $\sim$ 1$''$ resolution at the center
of NGC~1550 and good coverage of the group emission out to $\sim$ 200 kpc.
The data were telemetered in Very Faint mode, which can significantly
reduce the soft particle background. \asca\ grades 1, 5 and 7 were excluded.
We applied the CXC correction on charge transfer inefficiency (CTI).
The slow gain changes in ACIS CCDs I0-I3 were also corrected using the program
`corr\_tgain' by A. Vikhlinin
\footnote{http://cxc.harvard.edu/contrib/alexey/tgain/tgain.html}.
Known bad columns, hot pixels, and CCD node boundaries also were excluded.
We investigated the background light curves from chip S2. No background
flares were found in either observation.

\subsection{Background subtraction \& spectral fitting}

Background subtraction is critical for the analysis of extended sources.
In our analysis we used period D blank field background data (CTI corrected
by M. Markevitch\footnote{http://cxc.harvard.edu/contrib/maxim/bg/index.html}).
The particle background level was measured in PHA channels 2500-3000 ADU
for all CCDs. We found that the particle background levels were 4.8\% and
5.6\% higher for Obs 3186 and 3187 respectively than that of the period D
background data. This is within the uncertainties of the background files.
Thus, we increased the background normalization by 5.6\% and 4.8\% to fit
the particle background level at the time of the NGC~1550 observation
(see Markevitch et al. 2003). We also verified that the \rosat\ soft
sky background flux at the position of NGC~1550 matches quite well
with the average value of soft sky background in the \chandra\
background fields (within 1\%). A 10\% uncertainty of background is
included in the spectral fitting.

CALCARF and CALCRMF were used to generate response files.
The spatial non-uniformity of the CCD quantum efficiency (QE) was
included in ancillary response files (ARFs).
Two corrections were further made to the ACIS low energy quantum efficiency
(QE). The first corrects for the QE degradation, which increases with
time and is uniform in the detector plane. The second corrects the QE
by an empirical factor of 0.93
below 1.8 keV in the FI CCDs to improve the cross-calibration with the BI
CCDs\footnote{http://asc.harvard.edu/cal/Links/Acis/acis/Cal\_prods/qe/12\_01\_00/} 
(see Markevitch \& Vikhlinin 2001 for details).
Throughout the paper, we use the Galactic absorption of 1.14$\times$10$^{21}$
cm$^{-2}$. With this absorption value, the soft band spectra (0.6 - 1.0 keV)
can be well fit. To fit the NGC~1550 spectra in large regions, we generally
used the 0.7 - 6 keV energy band. In some regions, there is hard X-ray
excess in the spectrum of Obs 3186 but it only has a very small effect
on the spectral fitting.
The calibration files used correspond to CALDB 2.21 from the CXC. The
uncertainties quoted in this paper are 90\% confidence intervals
unless specified. The solar photospheric abundances of
Anders \& Grevesse (1989) are used in the spectral fits.

\subsection{X-ray morphology and surface brightness}

The two \chandra\ pointings of NGC~1550 were combined and the 0.5 -
3 keV image (background-subtracted and exposure-corrected) was produced.
The X-ray contours of the image superposed on the DSS II image are shown
in Fig. 1. All point sources are replaced by
surrounding averages. The image reveals some asymmetry in the center, but is
more relaxed than several other bright 1 keV groups, e.g., NGC 5044 (Buote et
al. 2003a) and NGC 507 (Kraft et al. 2003). The diffuse X-ray
emission can be traced to the edge of the \chandra\ field.
The X-ray surface brightness profile of NGC~1550 is shown in Fig. 2.
Beyond the central 40 kpc, the surface brightness is approximately
characterized by a power-law with a slope corresponding to $\beta$=0.46.
If the surface brightness profile is fit by a single $\beta$-model,
the central excess is quite significant and the fit cannot well describe
the measured profile at radii larger than 100 kpc (Fig. 2). Even a double
$\beta$-model cannot reproduce the excess in the central 1 kpc, but the fit
to radii larger than 1 kpc is much better than the single $\beta$-model fit.
The more diffuse component has a core radius of $\sim$ 26 kpc and $\beta$ =
0.48, which is consistent with the results based on \rosat\ observations
(HP00) and simulations by Muanwong et al. (2002), but smaller than the
medians of simulations by Dav$\acute{\rm e}$, Katz \& Weinberg (2002)
($\beta \sim$ 0.66) and TN01 ($\beta \sim$ 0.9). However, the limited
fitting range in the \chandra\ observations and the degeneracy between the
core radius and $\beta$ bias $\beta$ to small values.
Based on the \rosat\ All-Sky Survey data, we find the surface brightness
steepens beyond 200 kpc. The surface brightness can be characterized by a power-law
with a slope corresponding to $\beta$=0.70$^{+0.13}_{-0.12}$ between 200 kpc
and 450 kpc.

We also performed 2-D fits to the \chandra\ image of the NGC~1550 group
using Sherpa. The results agree with those of the 1-D fits.
No significant offset component was found. From small scales (e.g., 10 kpc)
to large scales (e.g., 200 kpc), the X-ray emission is somewhat elongated
in the E-W direction with ellipticities ranging from 0.1 to 0.2.

\subsection{Average temperature and abundances}

The integrated 0.72 - 6 keV spectra of Obs 3186 and 3187 within 200 kpc
were fit by a MEKAL model. Two pointings yield same results:
T=1.37$^{+0.02}_{-0.01}$ keV and abundance = 0.26$\pm$0.02 for Obs 3186,
and T=1.38$^{+0.01}_{-0.02}$ keV and abundance = 0.27$\pm$0.02 for Obs 3187.
Simultaneous fits yield: T=1.38$^{+0.01}_{-0.02}$ keV and abundance = 0.27$^{+0.01}_{-0.02}$
($\chi^{2}_{\nu}$=279.0/236). The Fe-L blend and S He-$\alpha$ lines can be
clearly seen in the integrated spectra (see Fig. 3). To determine the abundance of
individual elements, we also fit the spectrum with a VMEKAL model. Following
Finoguenov, Arnaud \& David (2001), we divide heavy elements into five groups
for fitting: Ne; Mg; Si; S and Ar; Ca, Fe, and Ni. The best-fit values
are: T=1.37$\pm$0.01 keV, Ne$<$0.11, Mg=0.08$^{+0.09}_{-0.04}$,
Si=0.28$^{+0.04}_{-0.05}$, S=0.41$^{+0.08}_{-0.07}$, and Fe=0.33$\pm$0.02
($\chi^{2}_{\nu}$=247.0/232). The integrated spectra, as well as the
best VMEKAL fits, are plotted in Fig. 3.

We use the empirical relation derived in Evrard, Metzler \& Navarro
(1996, EMN96 hereafter) to estimate the virial radius:
\begin{equation}  \label{rv}
r_{\rm vir} = 2.78  h_{0.7}^{-1}  (T/10 {\rm keV})^{1/2} (1+z)^{-3/2} {\rm Mpc}
\label{eq:rvir}
\end{equation}
This scaling relation may not apply for cool groups (e.g., Sanderson et al. 2003),
but can be used for comparison with other groups.
Thus, for NGC~1550, \rv = 1.01 Mpc.

\subsection{Radial temperature \& abundance profiles}

Although in X-rays, the NGC~1550 group is not perfectly azimuthally symmetric,
it is useful to derive radially averaged profiles of some physical properties
to compare with other groups and simulations. We took
the center for all annuli at R.A. = 04$^{h}$19$^{m}$37$^{s}$.9, decl. =
02\mydegree~24\myarcmin35\myarcsec, which is the peak of the X-ray distribution
and coincides with the optical galaxy center.
We required each annulus to contain 1500 - 2000 source counts from the two pointings.
Point sources are excluded. In the outermost bin (radius 9.5$'$ - 13.1$'$),
the group emission still contributes $\sim$ 60\% of the total counts in the 0.7
- 4 keV energy band. Each annulus was fit by a MEKAL model, with the temperature
and abundance as free parameters. The spectral fits from the two
pointings separately agree with each other very well in all annuli. Thus, we
made simultaneous fits to the two pointings. $\chi^{2}_{\nu}$ values
range from 0.7 to 1.17 (for 23 - 156 degrees of freedom).
The derived temperature and abundance profiles are shown in Fig. 4.
The temperature increases from $\sim 1$ keV at the center to $\sim 1.6$ keV
at $\sim 30$ - 100 kpc, then decreases to $\sim$ 1.0 keV at 200 kpc.
The heavy element abundance decreases from $\sim$ 0.5 solar at the
center to $\sim$ 0.1 solar at 200 kpc. The average abundance of 0.27
is typical for galaxy groups (Mulchaey \& Zabludoff 1998).

We also extracted the spectra of regions outside 200 kpc or 13.1$'$
(mostly from the S2 chip in Obs 3187). The \chandra\ observations only cover $\sim$
8\% of the projected area between radii of 200 - 380 kpc. The best-fit
temperature and abundance are 1.01$^{+0.10}_{-0.15}$ keV and
0.16$^{+0.15}_{-0.09}$ respectively. This implies that the temperature is either
flat or still decreasing beyond 200 kpc (Fig. 4).

In view of the strong line emission in the spectra of the NGC~1550 group,
we can constrain the radial abundance distribution of Si, S and Fe.
The radial profiles of Si, S and Fe abundances, shown in Fig. 5, are
obtained from an absorbed VMEKAL model fit. Abundances of Si, S and Fe
decrease with radius. Si/Fe is close to 1 in all bins. The abundances of
Ne and Mg are poorly constrained and are not shown.

\subsection{Total X-ray luminosity}

We estimate the total X-ray luminosity of the NGC~1550 group from the
spectral fits of the integrated spectra within 200 kpc.
The missing parts of the group (due to chip edges, chip gaps and point sources)
are included based on the measured surface brightness profile (double $\beta$-model
fit). Both pointings yield results within 1.5 \% and we average them to
obtain, the rest-frame 0.5 - 2 keV luminosity of 7.66$\times$10$^{42}$ erg s$^{-1}$
and bolometric luminosity of 1.65$\times$10$^{43}$ erg s$^{-1}$ within
200 kpc (or 0.20 \rv). If we extrapolate the surface brightness to
$r_{500}$ (or 650 kpc) and \rv (or 1.01 Mpc) based on the \rosat\
All-Sky Survey data, the bolometric luminosities are 2.0 and
2.3$\times$10$^{43}$ erg s$^{-1}$ respectively. For its temperature
(1.37 keV) and velocity dispersion ($\sim$ 300 km s$^{-1}$), the NGC~1550
group is among the X-ray brightest galaxy groups in the sky
(HP00; Mulchaey et al. 2003).

\subsection{Deprojection analysis}

Assuming spherical symmetry, we performed a spectral deprojection analysis to
obtain the 3D temperature and abundance profiles. Nine radial bins were
chosen to do the spectral deprojection. We used the standard `onion
peeling' technique (e.g., Sun et al. 2003). The projected emission measure
from the outer annuli on the inner annuli is estimated. The missing emission volume
in each annulus because of the chip edges, gaps and point sources is derived
by Monte Carlo simulations and included in the computation.
The deprojected temperatures and abundances are shown in Fig. 4. 

The electron density profile can be obtained by deprojecting the surface
brightness profile. This technique, starting in the largest annulus, converts
the observed surface brightness to electron density and then determines the
density at progressively smaller radii, after removing the projected emission from
larger radii. The projected flux contribution from regions beyond the
outermost bin is also subtracted based on the \rosat\ All-Sky Survey data.
A 0.5 - 3 keV background-subtracted and exposure-corrected surface brightness profile
of NGC~1550 is derived in 28 radial bins. With the knowledge of the group emissivity 
in these bins, we can derive electron densities. In the temperature
range appropriate to the NGC~1550 group ($\sim$ 1 keV), the X-ray emissivity
is sensitive to the heavy element abundance and temperature. Thus,
we need to model both the temperature and abundance profiles. The
deprojected temperature and abundance profiles can be well fit by fifth-order
polynomials. Repeated fitting 1000 times, we obtain the temperature and
abundance values in each radial bin, as well as $1\sigma$ uncertainties.
With these values, we can derive the 0.5 - 3 keV emissivity profile in the
28 bins from 1000 XSPEC simulations, as well as the corresponding uncertainties.
We can then use the standard `onion peeling' technique to convert the
surface brightness to electron density, where the conversion factor in
each bin is determined by the X-ray emissivity in this bin. The uncertainties
of the electron densities were obtained from the 1000 Monte Carlo simulations.

The derived electron density profile is shown in Fig. 6. It can be fit well by
a $\beta$-model if the innermost bin is excluded ($\chi^{2}_{\nu}$ = 28.2/24). The
small core radius (r$_{0}$=2.49$^{+0.63}_{-0.58}$ kpc) and $\beta$ (0.364$\pm$0.008)
imply that n$_{\rm e} \propto$ r$^{-1.09 \pm 0.02}$ between 10 kpc and 200 kpc.
This density profile is much flatter than those measured in clusters,
generally n$_{\rm e} \propto$ r$^{-2}$ beyond several core radii. With the
deprojected temperature and electron density profiles, we also derived the
cooling time and entropy profiles (defined as $S = kT/n_{\rm e}^{2/3}$) (Fig. 7).
The cooling time is less than 10$^{8}$ yr in
the very center and less than a Hubble time ($\sim$ 10$^{10}$ yr) within
the central 60 kpc. Thus, the decrease of temperature towards the
center within the central 40 kpc can be explained by radiative cooling.

\section{Optical properties of the NGC~1550 group}

NGC~1550 was shown to be a dominant galaxy of a group by Garcia (1993)
and Giuricin et al. (2000), but the listed information is very limited.
we studied the galaxy distribution using data extracted from NASA/IPAC
Extragalactic Database (NED) within 1 Mpc (the virial radius) of NGC~1550.
The velocity information
is complete to B $\sim$ 15.5 mag. There are 15 galaxies that have similar
redshifts to NGC~1550. We plot the distribution of these galaxies in
Fig. 8. These galaxies accumulate in a loose way that there is no concentration
at the core (e.g., 0.2 \rv). Using robust estimators of location and scale
(Beers, Flynn \& Gebhardt 1990), we obtain a mean velocity of
3694$^{+77}_{-66}$ km s$^{-1}$ and a velocity dispersion of
311$\pm$51 km s$^{-1}$. The mean velocity is essentially same as the
velocity of NGC~1550 (3714$\pm$23 km s$^{-1}$), which may imply the small
peculiar velocity of NGC~1550. The velocity dispersion of NGC~1550 implies
it has a low $\beta_{\rm spec}$ ($\mu$m$_{\rm p}\sigma_{\rm r}^{2}$/kT)
of $\sim$ 0.48, which is very similar to the other two well-studied 1 keV
groups NGC 2563 ($\beta_{\rm spec}$ = 0.52) and NGC 4325 ($\beta_{\rm spec}$
= 0.42) (Mushotzky et al. 2003, M03 hereafter). This value is also consistent
with the simulation results in Dav$\acute{\rm e}$ et al. (2002).
The $\beta_{\rm spec}$ of NGC~1550 is also similar to its $\beta_{\rm fit}$
within 200 kpc (0.41 - 0.48 depends on single or double $\beta$-model fits).

Within 0.5 \rv (or 33.1$'$) of NGC~1550, there are five bright group
members other than NGC 1550 (3 UGC galaxies and 2 CGCG galaxies).
At least three of them are brighter than R$_{\rm cD}$+2, where
R$_{\rm cD}$ is the R magnitude of the cD. Thus, the NGC~1550 group
is not a ``fossil group''.

The optical properties of the cD galaxy --- NGC~1550 are important
for us to understand the group properties. We have used the Danish 1.54m
telescope at La Silla, Chile, to obtain a (mosaic) B-band image (600 sec)
and an R-band image (450 sec) of NGC 1550. Standard optical data reduction
is followed. The optical light profile within 60 kpc can be well fit
by a de Vaucouleurs r$^{1/4}$ profile with a half light radius of 9.1 kpc.
There is $\sim$ 20\% excess beyond 35 kpc. The measured R-band magnitude
of NGC~1550 within 60 kpc is 10.97 mag after Galactic extinction correction
and K correction, which corresponds an R-band luminosity of 6.04 $\times$
10$^{10}$ L$_{\odot}$.
The galaxy is slightly elongated in the NE-SW direction and the
ellipticity is small (0.10 - 0.17). There is a dwarf galaxy
concentration within 5$'$ of NGC~1550. NGC~1550 has a normal absorption
line late-type galaxy spectrum (from the CfA database).

\section{Mass profile \& M-T relation}

\subsection{X-ray Gas Mass and Total Gravitating Mass}

The gas density and temperature profiles can be used to derive the total
gravitational mass profile under the assumption of hydrostatic
equilibrium:
\begin{equation}  \label{M}
M (<r) = - \frac{k T(r) r}{\mu m_p G} (\frac{d \log n_e(r)}{d \log r} + \frac{d \log T(r)}{d \log r})
\label{eq:mass}
\end{equation}

To overcome the difficulty of deriving the logarithmic derivative of
the gas temperature profile, we fit the observed temperature data to a
fifth order polynomial since the temperature profile was not well fit
by any simple function. To constrain the temperature gradient at
$\sim200$ kpc, we included the measured temperature value between 200
and 380 kpc from the small region in that annulus (see $\S$2.5) and
increased its uncertainty by a factor of two. The fit is very good
($\chi^2_{\nu}$ = 8.7/5). To derive the
temperature and temperature gradients at the grid points corresponding
to the more accurately measured gas density, we performed Monte
Carlo simulations. In each run, the temperatures are scattered with
the measured uncertainties around the measured values. Each
simulated temperature profile is then fit by a fifth-order
polynomial. Repeating this procedure 1000 times, we obtain the
temperature and temperature gradient in each radial bin, as well as
$1\sigma$ uncertainties. The electron density profile is well fit by a
$\beta$-model except for the central point (Fig.~6) and a similar
method is applied to derive the density gradient and its
uncertainties. Thus, through a series of Monte Carlo simulations, we
derive the total mass profile (eq.~\ref{eq:mass}) outside the central
2~kpc. The gas mass profile can be derived from the deprojected
density and abundance profiles.

Both the total mass and gas mass profiles are shown in Fig. 9, along
with the gas mass fraction profile. To explore the gas fraction at
even larger radii, we estimate the total mass and the gas mass to
about $\sim300$ kpc based on the S2 chip temperature and the surface
brightness measured from the \rosat\ All-Sky Survey data. The estimates
are also plotted in in Fig. 9.
The gas mass fraction increases rapidly with radius to $\sim 0.07$ at
200~kpc. Beyond 200 kpc, uncertainties remain large and require
additional data to define its behaviour.

\subsection{Stellar Contribution to the Baryon Fraction and
the Mass-to-Light Ratio}

Since much of the baryon matter in the core of the NGC~1550 group
arises from stellar matter in galaxies, we have estimated the stellar
light profile to $\sim$ 300 kpc. Most of stellar light within 200~kpc
is contributed by NGC~1550, while the contributions from other member
galaxies become almost comparable to that of NGC~1550 at 300 kpc.
Since the light profile of NGC~1550 can be well fit by a de
Vaucouleurs profile, we use the deprojection table (Young 1976)
assuming an R-band mass-to-light ratio of $M/L = 5
M_{\odot}/L_{\odot}$ (Kauffman et al. 2003).  With these assumptions,
we find a total stellar mass of NGC~1550 within 60~kpc of $3.0 \times
10^{11} M_{\odot}$. We also can estimate the total stellar mass of
NGC~1550 from its K-band luminosity by assuming a K-band mass-to-light
ratio of $\sim1$ (Cole et al. 2001). This yields a total stellar mass
of $\sim2.4\times10^{11} M_{\odot}$, consistent with our R-band estimate
under the uncertainties of both mass-to-light ratios.
We also include the optical light from other group member galaxies and half
that of dwarf galaxies (in projection without measured velocities) in the
light profile at their projected radii.  Since we do not have R-band
CCD data for galaxies beyond the central 190 kpc, we derive their
stellar mass contributions from their K-band magnitudes
measured from the Two Micron All Sky Survey (2MASS). 
We increase these measured values by 20\% to derive the total galaxy
light (Kochanek et al. 2001) along with an assumed K-band mass-to-light
ratio of 1.

The derived stellar mass profile, as well as the baryon fraction
profile, are shown in Fig.~9. As expected, baryonic matter dominates in the core of
NGC~1550. The baryon fraction reaches $\sim0.1$ at 190~kpc. The
R-band mass-to-light ratio of the group increases with radius and
reaches $155\pm35$ at 190~kpc. A similar value of M/L is found in the
V band. This value is similar to those found at $\sim$ 0.2 \rv of
clusters and groups (e.g., David et al. 1995; Pratt \& Arnaud 2003).
Beyond 190 kpc, there are some bright member galaxies (7 UGC galalxies
+ others) and M/L is not expected to change much beyond 0.2 \rv (e.g., Pratt \&
Arnaud 2003). Thus, The NGC~1550 group is not an OLEG.

\subsection{Comparison of the Total Mass Distribution to Models and
the $M-T$ Relation}

Using the derived total mass distribution, we can test numerical
models of the growth of dark matter halos (e.g., Navarro, Frenk, \&
White 1997; Moore et al. 1998). However, the two parameters in these
profiles, $\delta_{\rm c}$ (central overdensity) and $r_{\rm s}$
(characteristic radius), are highly degenerate in fitting the accumulating
mass profile. This, in combination
with the limited range of the measured mass profile, makes it
impossible to derive a global dark matter model for the NGC~1550
group.  Nevertheless, we fit the total mass profile within 200~kpc to
obtain an estimate of the concentration parameter. For an NFW profile,
we obtain best-fit parameters of $\delta_{\rm c} = 1.10 \times 10^{5}$
and $r_{\rm s} = 41.8$~kpc ($\chi^{2}_{\nu}$ = 38.7/25). The Moore
profile better describes the mass profile of NGC~1550 ($\chi^{2}_{\nu}$
= 13.5/25), especially within the central 10~kpc with $\delta_{\rm c}$
= 8.5 (3.7 - 17.5) $\times$ 10$^{3}$ and r$_{\rm s}$ = 105 (70 - 170)
kpc (note the large degeneracy between the two fitted parameters).
The corresponding concentration is $\sim$ 6.2, which implies a halo
mass of $\sim$ 10$^{13}$ M$_{\odot}$. It would be most interesting
to compare the theoretical models to more accurate mass profile
extending over a larger range of radii to critically test their
similarities and differences in low mass systems.

Allen, Schmidt \& Fabian (2001) derived an $M-T$ relation for hot
clusters within $r_{2500}$. If cool groups also follow this relation,
we expect $M_{2500} = (2.7\pm0.3) \times 10^{13} M_{\odot} (T / 1.37
keV)^{1.5}$ for NGC~1550. Since, for NGC~1550, $r_{2500}$ is slightly
beyond 200~kpc, we extrapolated the mass profile to $r_{2500}$. The
Moore profile fit predicts $r_{2500}$ = 208 - 237 kpc and $M_{2500}$
= 1.5 - 1.8 $\times 10^{13}$ M$_{\odot}$, which is smaller than the
value expected from the $M-T$ relations of hot clusters.

\section{Discussion}

\subsection{The temperature profiles of 1 keV galaxy groups}

NGC~1550 exhibits a unique temperature profile and shows similarities
to other cool groups with accurate measured profiles. In Fig.~10, we
show deprojected temperature profiles of the four groups NGC~1550, NGC
2563, NGC 4325 and NGC 5044 with accurate temperature profiles
measured with \chandra\ and \xmm\ (M03; Buote et al. 2003a, see David
et al. 1994 for the outer profile of NGC~5044 measured with ROSAT). To
compare different groups, we rescaled the radii by the virial radius
for each group (computed from eq.~\ref{eq:rvir}) and rescaled the
temperatures by the mean group temperature. Remarkably, the
temperature profiles of NGC~1550, NGC 2563 and NGC 5044 are very
similar with a temperature peak or plateau between \rv=0.04 and 0.08,
and temperature declines, with similar slopes, beyond $\sim0.08$
\rv.  NGC 4325 seems to have a rather flatter temperature profile, but
still shows a decline beyond $\sim0.08$ \rv. 

The \rosat\ and \asca\ temperature profiles of $\sim1$~keV groups,
although with much poorer statistical precision show similar average
features. The four $\sim1$~keV groups studied by HP00 with the best
measured temperature profiles, NGC 741, NGC 4065, NGC 4761, and NGC
5846, all show similar temperature declines beyond $\sim0.1$ \rv.
Similar temperature profile of NGC 4761 (or HCG 62) was also found
by Mulchaey \& Zabludoff (1998) and Buote (2000).
Similarly, nine galaxy groups observed with ASCA also generally show
temperature declines at large radii (from 0.1 - 0.2 \rv) (see
Finoguenov et al. 2002). However, all these temperature profiles have
large uncertainties and insufficient angular resolution to directly compare with
the precisely measured \chandra\ and \xmm\ profiles shown in Fig.~10.

In summary, the current knowledge of 1~keV galaxy groups suggests that
at least some of them have a rather common temperature profile with
temperatures declining beyond $\sim0.08$ \rv, and the temperature at
0.2 \rv is $\sim70$\% of the temperature at 0.05 \rv. However, there
also seem to exist a class of 1~keV galaxy groups with very flat
temperature profiles including NGC~2300 (Davis et al. 1996) and
NGC~4325 (M03).  This may reflect different roles of non-gravitational
processes (including cD heating) or different formation histories in
different groups. More high-quality temperature profiles from \chandra\
and \xmm\ are needed to fully understand the origin of temperature
structure of cool groups.

\subsection{Comparison to temperature profiles of hot clusters}

Hot galaxy clusters show some of the same temperature profile
characteristics as seen for NGC~1550, but with an important difference.
Based on the analysis of \asca\ data, Markevitch et al. (1998;
M98 hereafter) argued that hot clusters exhibit a universal temperature
profile, with the temperature at 0.5 \rv declining to only $\sim50$\%
of the temperature at 0.1 \rv. De Grandi \& Molendi (2002; DM02 hereafter)
using \sax\ data found good agreement with M98 beyond 0.2 \rv and a flat
profile from the cooling radius to 0.2 \rv. Despite the agreement
between DM02 and M98, the behavior of cluster temperature profiles
beyond 0.2 \rv remains controversial with cluster showing flat
profiles to 0.4 - 0.7 \rv (e.g., A2390, A1795, A1835 and A2163)
(Allen, Ettori \& Fabian 2001; Arnaud et al. 2001; Pratt, Arnaud
\& Aghanim 2001; Majerowicz, Neumann \& Reiprich 2002).
However, some clusters do show temperature declines at large radii
(e.g., A1413, Pratt \& Arnaud 2002). In summary, recent observations support flat
temperature profiles of hot clusters beyond the cooling radius to at
least 0.2 \rv, while the behaviour beyond 0.2 \rv is not definitively
determined.

Despite the uncertainty in cluster temperature profiles are large
radii, their behavior at moderate radii contrasts strongly with that
of the 1~keV galaxy groups shown in Fig.~10, in which significant
temperature declines are found between 0.1 \rv and 0.2 \rv. To
quantitatively compare clusters and groups, we fit the temperature
profiles of 1~keV groups at large radii (beyond 0.07 \rv) with a polytropic
model ($T/<T> \propto (r/$\rv$)^{-\mu}$) to measure the slope of
the temperature decline. Combining all four profiles from Fig.~10,
the best-fit is $\mu = 0.37\pm0.03$ ($\chi^{2}_{\nu}$ = 29.7/11).
If we omit the rather flat temperature profile of NGC 4325, the results
change little: $\mu = 0.40\pm0.03$ ($\chi^{2}_{\nu}$ = 18.1/8).
While the polytropic index we find for groups is remarkably similar
to that found by DM02 for cooling-flow clusters, the temperature declines
in groups occur at much smaller radii -- 0.07 -0.25 \rv for groups
compared to 0.2-0.6 \rv as found by DM02 for hot clusters. If we
approximate the gas density profile in this range as $n(r)\propto$
$r^{-\nu}$ with $\nu = 3/2$ or 2, expected if the gas density profiles
follow a $\beta$-model with $\beta = 1/2$ or 2/3 and the pressure is
given by $p \propto n^{-\gamma}$, where $\gamma = \mu/\nu + 1$, then we
find $\gamma$ = 1.26 - 1.19 (with an uncertainty of 0.02. Thus,
$\gamma$ lies between the isothermal index of 1 and the adiabatic index
of 5/3, in agreement with the value found by M98, $\gamma$ =
1.24$^{+0.20}_{-0.12}$ for hot clusters, but, again, at larger radii.

The observed difference in the radius at which groups and clusters
show the radial temperature decline can be alleviated if the virial
radius of cool groups is smaller than that derived from the empirical
relation of EMN96 (see eq.~\ref{eq:rvir}).  The Birmingham-CfA cluster
scaling project reveals that the measured virial radii of clusters and
groups are smaller than those expected from self-similar scaling
relations, especially for cool groups (Sanderson et al. 2003). For
groups like NGC~1550, their result shows that the true virial radius
may be as small as 60\% of the value derived using the empirical
relation by EMN96. In fact, if we applies the best-fit Moore profile in
$\S$4.3, the derived virial radius of the NGC~1550 group is $\sim$ 650
kpc --- $\sim$ 65\% of the value derived using the simple scale relation.
Nevertheless, this cannot completely explain the observed difference.

\subsection{Comparison with temperature profiles from simulations}

The temperature profiles of these $\sim$ 1 keV groups at radii larger
than 0.05 \rv can be compared to simulations.
Most simulations are for hot clusters, while the ones dedicated to cool groups
are rather scarce. The observed 1 keV group temperature profiles (see Fig. 10) agree
with the ``Universal temperature profile'' of galaxy clusters from adiabatic
simulations (Loken et al. 2003). However, these simulations are aimed
at hot and massive clusters where the non-gravitational processes are not
expected to be significant. ``Entropy floor'' simulations (TN01;
Borgani et al. 2002; Babul et al. 2002) can reproduce the observed temperature
profiles, but no entropy floor is found in the NGC~1550 group (Fig. 7).
Muanwong et al. (2002) produced adiabatic, radiative and pre-heating
simulations of galaxy groups. Their predicted temperature profiles are too
flat between 0.06 \rv and 0.2 \rv, inconsistent with the majority of
observations. Dav$\acute{\rm e}$ et al. (2002) include cooling and star
formation, but no substantial non-gravitational heating in their simulations
of galaxy groups. The resulting temperature profile is consistent with the
observations. Temperature and entropy profiles to $\sim$ 0.5 \rv will
be able to put tighter constraint on simulations in the future.

The inner (\lsim 0.05 \rv) temperature profiles of 1 keV groups (Fig. 10)
are also similar. None of simulations describe the temperature profiles
in this region well. This may require high resolution simulations
that carefully handle the radiative cooling and a better understanding
of the cooling flow physics.

\subsection{The role of cD heating}

Since the temperature peak or plateau of these 1 keV groups lies at relatively
small radii (within 30 - 110 kpc), feedback from the cD galaxy should not
be ignored. The high temperature plateau at 30 - 110 kpc in the NGC~1550 group
may be caused by heating from the central galaxy. If we assume an energy input
of 0.2 keV / particle from NGC 1550, over the lifetime of the group an energy input of at
least 10$^{59}$ ergs is needed. This is $\sim$ 2 orders of magnitude higher than
the current radio output of NGC~1550 integrated over 10 Gyr. However,
NGC~1550 may have episodic activity related to the cooling flow (Sun et
al. 2003 and references therein). Galactic winds also can heat the ICM/IGM significantly.
Based on the simulations by David, Forman \& Jones (1991), the galactic wind
of NGC~1550 can provide up to an order of magnitude more energy than
required. Thus, NGC~1550 may play an important role in shaping the
inner temperature profile either during its radio active periods or
by its galactic wind. Thus, the temperature ``bump'' seen in other
groups also can be produced by the heating of the cD galaxies.

\subsection{The entropy profiles of 1 keV groups}

One scenario to explain the entropy excess in the center of galaxy groups is
that low-entropy gas in the central region cools, condenses into dense and
cold objects, and the high-entropy gas flows in to fill the volume.
Voit \& Bryan (2001) argued that only gas with entropy larger than 100 - 150
keV cm$^{2}$ has a long enough cooling time to survive in cluster/group
cores. However, we do not observe a flat entropy profile in the center of
NGC~1550, which questions the validity of simulations that invoke an
entropy-floor, e.g., TN01, Babul et al. (2002). This is consistent with
the recent systematic study by Ponman, Sanderson \& Finoguenov (2003).

The temperature decline from $\sim$ 0.08 \rv makes the entropy profile
flatten at large radii, $S(r) \propto r^{0.9}$ between 20 kpc and
80 kpc, while $S(r) \propto r^{0.2}$ between 80 kpc and 200 kpc (Fig. 7). A similar
phenomenon was found by Finoguenov et al. (2002) in several other galaxy
groups. If we adopt the S2 chip temperature at $\sim$ 300 kpc and the steepening
density profile from the \rosat\ All-Sky Survey data, the entropy profile
steepens again beyond 200 kpc. However, better measurements beyond 0.2 \rv
are needed to constrain the entropy profile at large radii.

At 0.1 \rv (101 kpc), the entropy is $\sim$ 120 keV cm$^{2}$, which
is consistent with the value of the entropy floor proposed by Ponman et
al. (1999) and Lloyd-Davies et al. (2000). However, M03
derived different entropy values at 0.1 \rv for two other
1 keV galaxy groups, NGC 4325 ($\sim$ 100 keV cm$^{2}$) and NGC 2563
($\sim$ 300 keV cm$^{2}$). Furthermore, Ponman et al. (2003) studied a
sample of 66 virialised systems. Their results prefer a slope in S(T)
which is significantly shallower than the self-similar relation S $\propto$
T rather than a fixed entropy floor at the core. They suggest that
entropy scales as S $\propto$ T$^{\sim 0.65}$.
Pratt \& Arnaud (2003) find that the entropy profiles of the 2 keV cluster
A1983 and the 7 keV cluster A1413 follow that relation.
We plot the scaled entropy profiles ((1+z)$^{2}T^{-0.65}S$) of three
1 keV groups and three 2 - 3 keV groups in Fig. 11. The entropy profiles
of the NGC 2563 and the NGC 4325 groups are from M03. The entropy profile
of Abell 1983 (T=2.1 keV) is from Pratt \& Arnaud (2003). The entropy
profiles of the ESO 3060170 group (T=2.70 keV) and the ESO 5520200 group
(T=2.10 keV) are from our work (Sun et al. in prep.). The scaled entropy profiles
of 2 - 3 keV groups align with each other better than those of 1 keV groups.
The scaled entropy profiles of these 1 keV groups flatten between $\sim$ 0.1
and 0.2 \rv, but with different values spanning almost a factor of three.
All these imply that the entropy profiles of 1 keV groups show larger
scattering than those of hotter groups, which reinforces the idea
that there is no fixed entropy floor for 1 keV groups. This reflects
larger roles of non-gravitational processes in 1 keV groups than hotter
systems and implies varied pre-heating levels.

\subsection{L$_{\rm x}$ - T \& L$_{\rm x}$ - $\sigma_{\rm r}$ relations}

Despite the known large scatter in the L$_{\rm x}$ - T relation,
NGC~1550 is very X-ray luminous for its X-ray gas temperature.
The NGC~1550 group is at least 5 times brighter than other groups in
the same temperature range (Mulchaey \& Zabludoff 1998).
If we compare it with galaxy groups in HP00,
it is $\sim$ 2 times brighter than any groups in the same
temperature range. Consistent with its somewhat extreme position
in the L$_{\rm x}$ - T relation, NGC~1550 is 3 times brighter than
the best-fit X-ray luminosity - galaxy velocity dispersion (L$_{\rm x}$
- $\sigma_{\rm r}$) in Mulchaey \& Zabludoff (1998) and at least
5 times brighter than the best-fit in HP00. We caution that the current
L$_{\rm x}$ - $\sigma_{\rm r}$ relations is not well constrained for cool
groups and there is large scatter.

\subsection{The abundances of heavy elements}

The abundance ratios of heavy elements give us clues to SN enrichment.
The global Si/Fe ratio is $\lsim$1, which agrees with that
found for galaxy groups (Finoguenov, David, \& Ponman 2000) and the value from recent
\xmm\ and \chandra\ observations of the NGC 5044 group (Buote et al. 2003b).
This ratio implies that $\sim$ 80\% of the iron is produced by SN Ia based on supernova
nucleosynthesis models (e.g., summarized by Gibson, Loewenstein \& Mushotzky
1997). This small ratio generally found in groups led Finoguenov et al.
(2000) to conclude that SN Ia play a larger role in the enrichment of groups
compared with that in clusters. S is relatively enriched in the group,
especially around the center. We need more precise and robust measurement
of Si and S abundances to better understand SN enrichment in this group.

\section{Conclusion}

We have presented an analysis of the \chandra\ observation of the NGC~1550
group. The derived physical properties of the NGC~1550 group are
summarized in Table 1. The main conclusion of our study are:

1. The NGC~1550 group is among the brightest X-ray groups
(L$_{\rm bol} \sim$ 1.65$\times$10$^{43}$ erg s$^{-1}$
within 200 kpc, or 0.2 \rv) in its temperature (1.37 keV) and
velocity dispersion ($\sim$ 300 km s$^{-1}$) range.

2. Like two other $\sim$ 1 keV groups (NGC 5044 and NGC 2563), the
temperature profile of the NGC~1550 group declines beyond $\sim$ 0.08
\rv and the declines are remarkably similar. The slope of
the decline is similar to that of the ``Universal temperature profile''
from adiabatic simulations (Loken et al. 2003), and also similar to
that of hot clusters (M98 and DM02). However, the cluster temperature
decline occurs at much larger radii (0.2 - 0.6 \rv). This difference may
reflect the roles of non-gravitational processes in ICM evolution.

3. The NGC~1550 group shows no isentropic core in its entropy profile, in
contradiction to
`entropy-floor' simulations. Because of the temperature decline beyond $\sim$
0.08 \rv, the entropy profile flattens from $\sim$ 0.08 \rv to
0.2 \rv. The entropy profiles of 1 keV groups within 0.2 \rv vary and
do not follow the $S \propto T^{\sim 0.65}$ scaling as well as hotter systems.
This may imply that the effects of non-gravitational processes vary for different
systems.

4. The stellar, gas and total mass profiles of the NGC~1550 group are derived
and show that the gas fraction and baryon fraction reach 0.07 and 0.1 respectively
at 190 kpc. The R-band mass-to-light ratio is 155$\pm$35 at 190 kpc.
The NGC~1550 group is not a ``fossil group'' or an OLEG.
The total mass profile can be better fit by Moore profile.

5. The Si/Fe ratio implies that SN Ia play a larger role for the enrichment
of the group than SN II. Si, S and Fe all enrich at the center compared to
the outer regions.

\acknowledgments

The results presented here are made possible by the successful effort of the
entire \emph{Chandra} team to build, launch, and operate the observatory.
This research has also made use of the NASA/IPAC Extragalactic Database (NED) which
is operated by the Jet Propulsion Laboratory, California Institute of Technology,
under contract with the National Aeronautics and Space Administration.
We thank the anonymous referee for prompt and valuable comments.
We thank T. Beers for his biweight estimators of location and scale.
We acknowledge helpful discussions with E. Churazov and M. Markevitch.
We acknowledge support from the Smithsonian Institution and NASA contracts
NAS8-38248 and NAS8-39073.

\vspace{-1cm}
\begin{table}
\begin{center}
\caption{The physical properties of the NGC~1550 group}
\vspace{0.3cm}
{\small
\begin{tabular}{c|c|c|c|c|c|c|c|c|c} \hline \hline
$\bar{\rm v}$ & $\sigma$ & N$_{\rm gal}^{\rm a}$ & Distance$^{\rm b}$ & N$_{\rm H}$ & T$^{c}$ & Si$^{c}$ & Fe$^{c}$ & Log L$_{\rm x}^{d}$ & Log (L$_{\rm R}$/L$_{\odot}$)$^{\rm e}$ \\
(km s$^{-1}$) & (km s$^{-1}$) & & (Mpc) & (10$^{21}$ cm$^{-2}$) & (keV) & (solar) & (solar) & (ergs s$^{-1}$) & \\ \hline
3694$^{+77}_{-66}$ & 311$\pm$51 & 16 & 53.7 & 1.14 & 1.37$\pm$0.01 & 0.28$^{+0.04}_{-0.05}$ & 0.33$\pm$0.02 & 43.22 & 10.78 \\
\hline \hline
\end{tabular}}
\begin{flushleft}
\leftskip 35pt
$^{\rm a}$ The velocity information is from NED.\\
$^{\rm b}$ For h$_{0}=0.7$\\
$^{\rm c}$ The emission-weighted temperature obtained from the VMEKAL model fit to the global spectrum\\
$^{\rm d}$ The bolometric luminosity within 200 kpc radius\\
$^{\rm e}$ The integrated R-band light within 60 kpc of NGC 1550 \\
\end{flushleft}
\end{center}
\end{table}
\clearpage

\begin{figure*}
\vspace{-3.5cm}
  \centerline{\includegraphics[height=0.9\linewidth,angle=270]{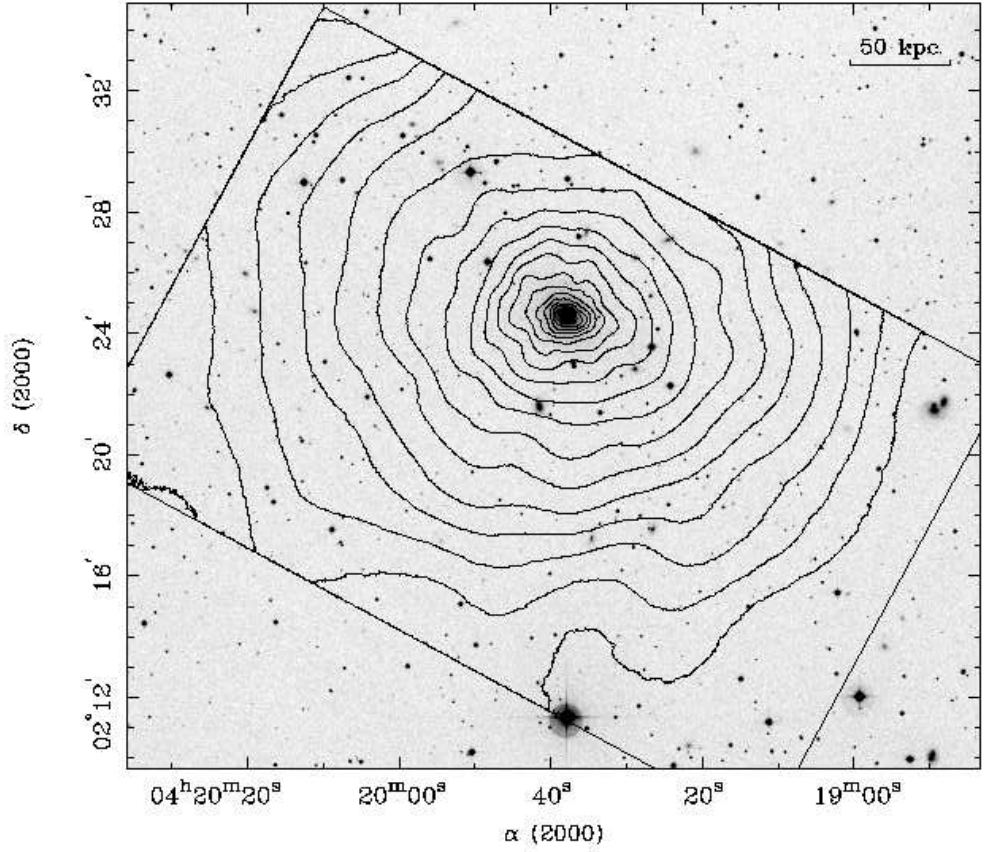}}
  \caption{\chandra\ 0.5 - 3 keV contours of the NGC~1550 group emission
(two pointings combined) superposed on DSS II image. The X-ray image
was background-subtracted and exposure-corrected. Point sources were
replaced by averages of surrounding diffuse emission. The X-ray image was
then adaptively smoothed. Only the emission in ACIS-I chips is shown.
The contours levels increase by a factor of $\sqrt{2}$ from the outermost one
(1.12$\times$10$^{-3}$ cts s$^{-1}$ arcmin$^{-2}$) towards
the center. The chip edges of the combined two ACIS-I fields are also shown.
NGC~1550 dominates the optical light in the field.
    \label{fig:img:smo}}
\end{figure*}
\clearpage

\begin{figure*}
\vspace{-2cm}
  \centerline{\includegraphics[height=0.6\linewidth,angle=270]{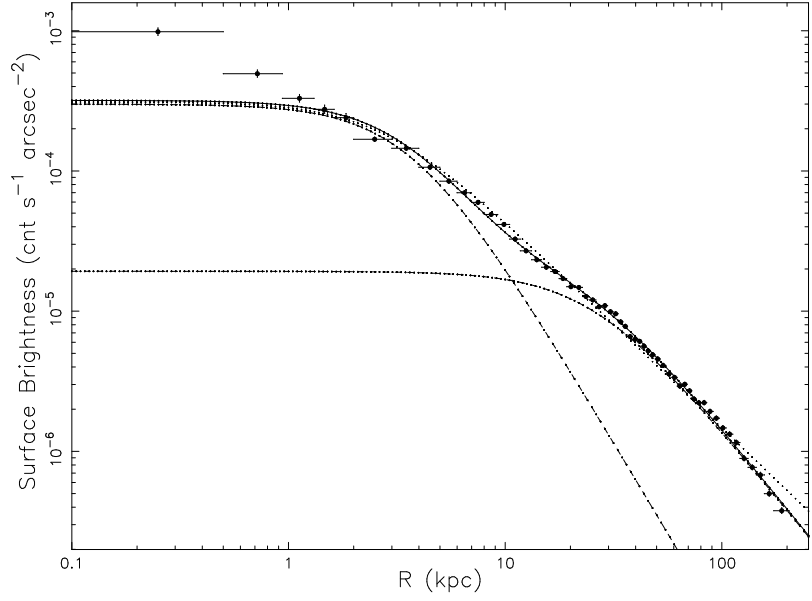}}
  \caption{The radial surface brightness profile of the NGC~1550 group measured
in the 0.5 - 3 keV band. It is fit with a single $\beta$-model
($\chi^{2}_{\nu}$ = 799.6/52; dotted line) and a double $\beta$-model
($\chi^{2}_{\nu}$ = 325.5/50; solid line with each component as a dotted-dashed
line). Neither model can fit the central peak, which requires an
unrealistically large $\beta$ if it is fit by a $\beta$-model.
   \label{fig:img:smo}}
\end{figure*}

\begin{figure*}
\vspace{-2cm}
  \centerline{\includegraphics[height=0.7\linewidth,angle=270]{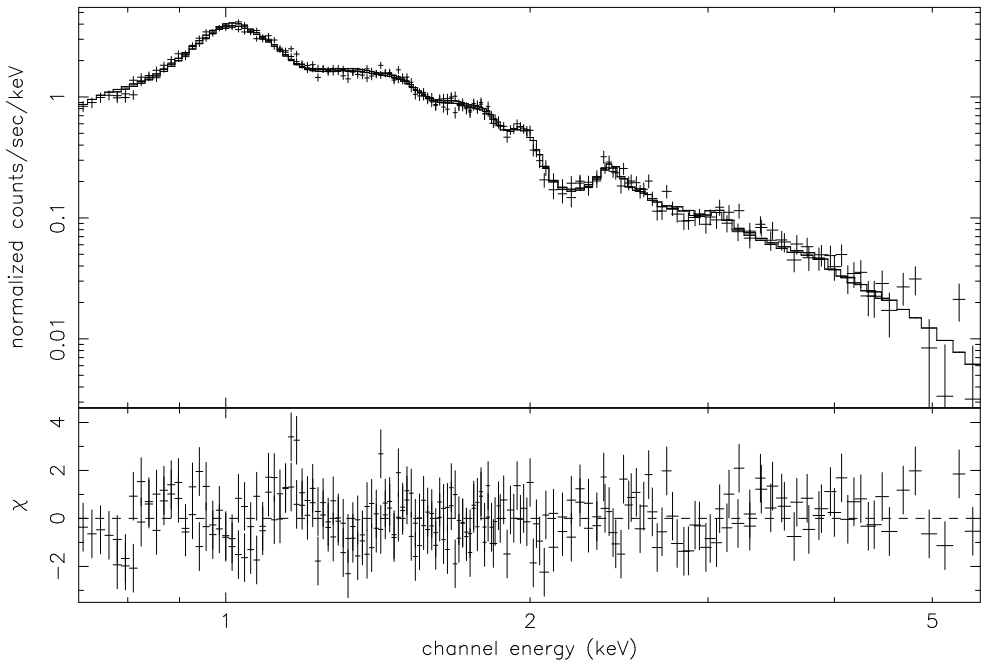}}
  \caption{The global spectra of NGC~1550 from two ACIS-I pointings, shown
with the best-fit VMEKAL model and residuals. The fit is good
($\chi^{2}_{\nu}$=247.0/232) and line emission from Fe-L blend (1.0 - 1.3
keV) and S He-$\alpha$ ($\sim$ 2.4 keV) is prominent.
   \label{fig:img:smo}}
\end{figure*}
\clearpage

\begin{figure*}
  \centerline{\includegraphics[height=0.8\linewidth,angle=270]{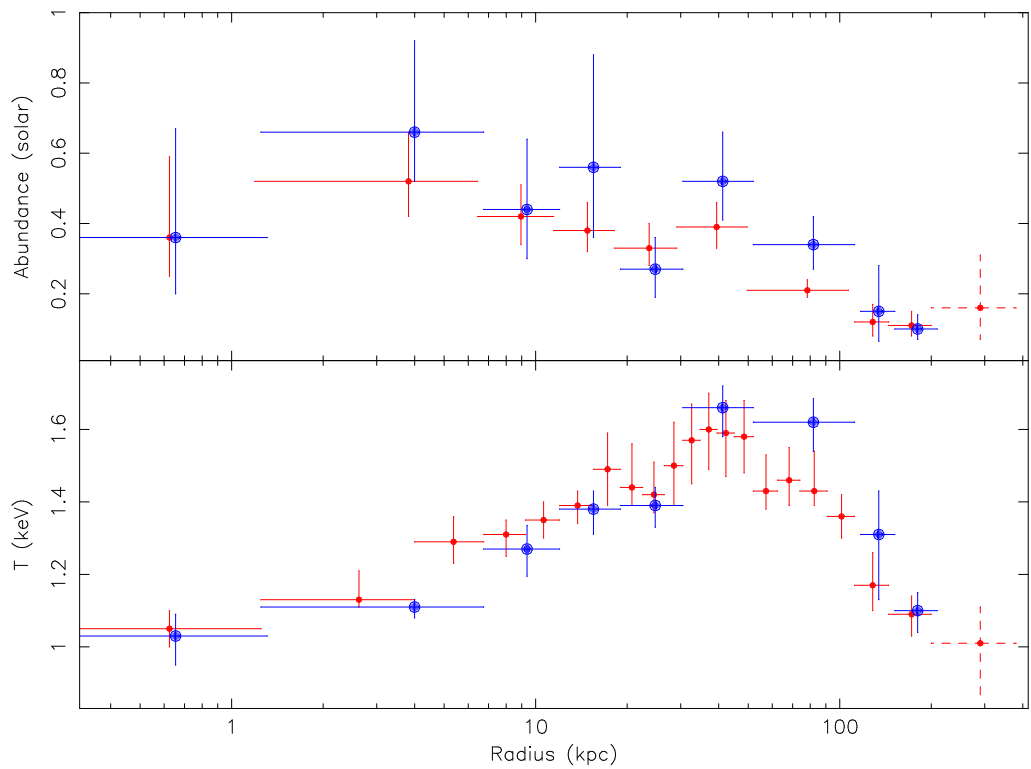}}
  \caption{Temperature and abundance profiles. The small red and large blue points
represent the projected and the deprojected values respectively. \chandra\ only
covers a very small region between 200 and 380 kpc (see text). The temperature
and abundance measured in that region are marked by points with dashed lines.
For deprojected values in the same bin as the projected values, we slightly
shift the deprojected values along the X axis for clarity.
The uncertainties are 90 \% confidence level values.
    \label{fig:img:smo}}
\end{figure*}
\clearpage

\begin{figure*}
  \centerline{\includegraphics[height=0.8\linewidth,angle=270]{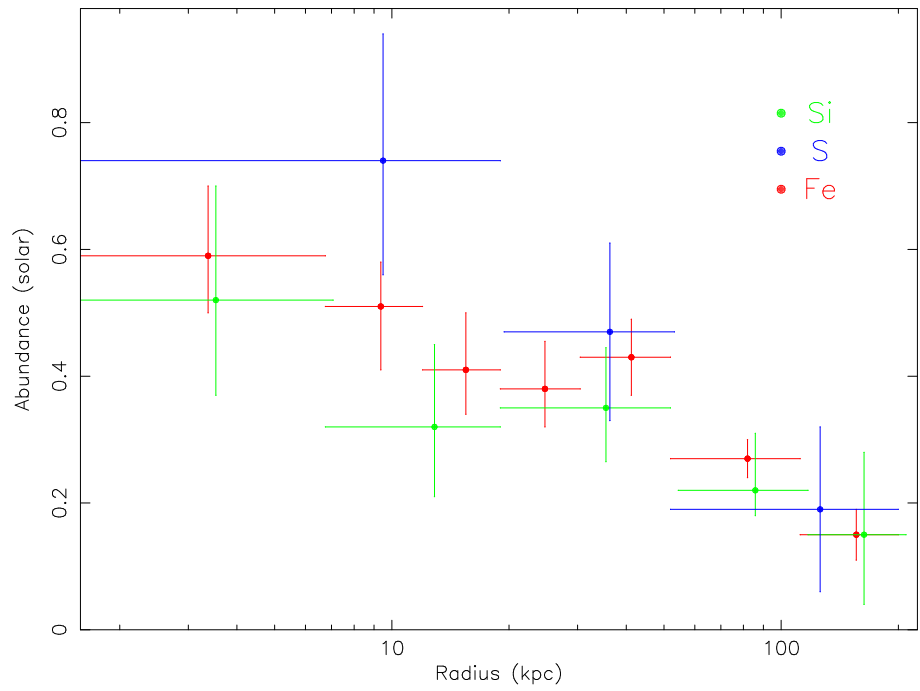}}
  \caption{The radial profiles of Si, S and Fe abundances. All profiles peak
at the center and S is specially enriched in the central 20 kpc.
    \label{fig:img:smo}}
\end{figure*}
\clearpage

\begin{figure*}
  \centerline{\includegraphics[height=1.0\linewidth,angle=270]{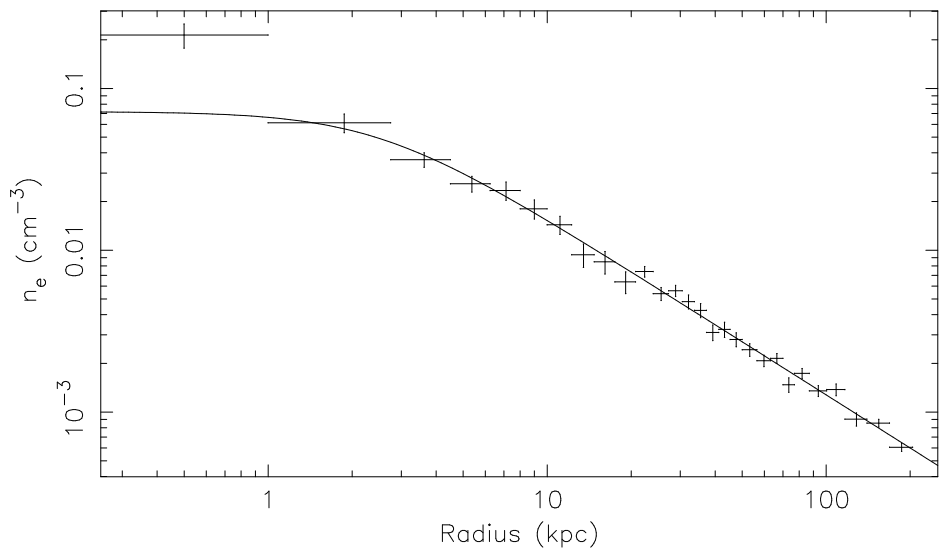}}
  \caption{The electron density profile obtained from the deprojection analysis
(1 $\sigma$ uncertainty). The solid line is the best-fit $\beta$ model to
all bins excluding the innermost one ($\chi^{2}_{\nu}$ = 28.2/24). Aside from
the central bin, the gas density distribution is well described by a single
$\beta$-model with $\beta$=0.364$\pm$0.008 and r$_{0}$=2.49$^{+0.63}_{-0.58}$
kpc.
    \label{fig:img:smo}}
\end{figure*}
\clearpage

\begin{figure*}
  \centerline{\includegraphics[height=1.0\linewidth,angle=270]{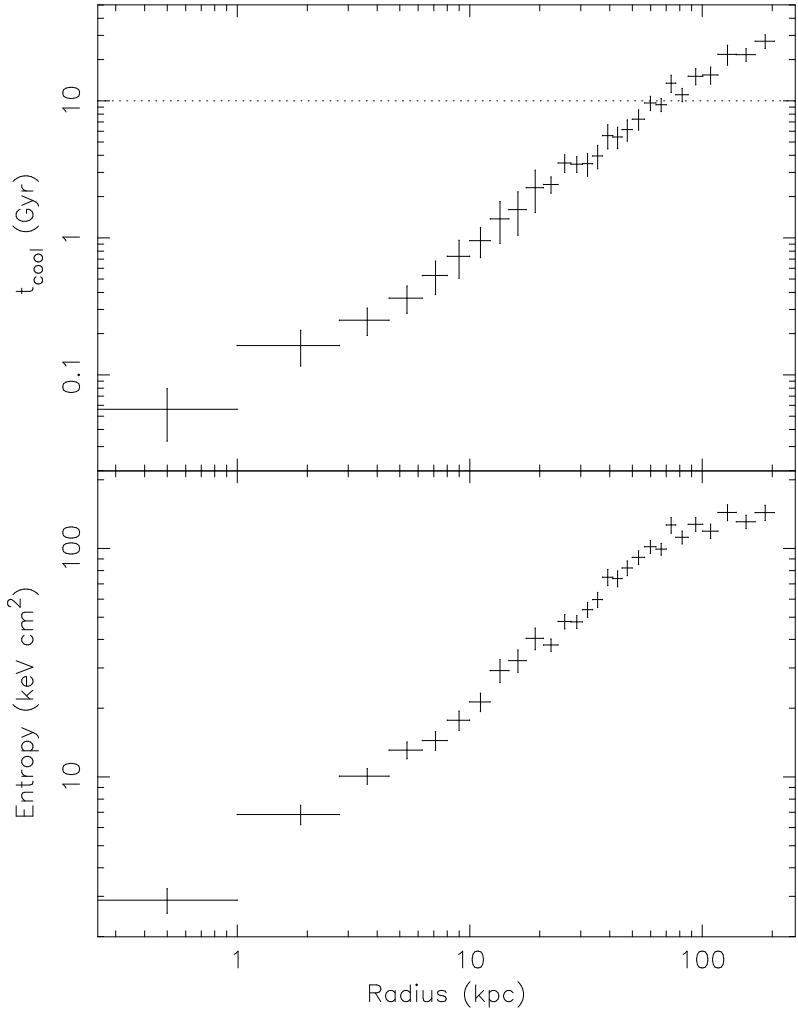}}
  \caption{Cooling time and entropy profiles, derived from electron
density profile and the best-fit to the deprojected temperature profile.
The gas cooling time profile shows that everywhere within $\sim$ 60 kpc,
the gas cooling time is less than a Hubble time. The entropy distribution
shows no evidence of an entropy floor, as suggested in some pre-heating
models. The entropy profile flattens between 80 and 200 kpc.
    \label{fig:img:smo}}
\end{figure*}
\clearpage

\begin{figure*}
  \centerline{\includegraphics[height=0.9\linewidth,angle=270]{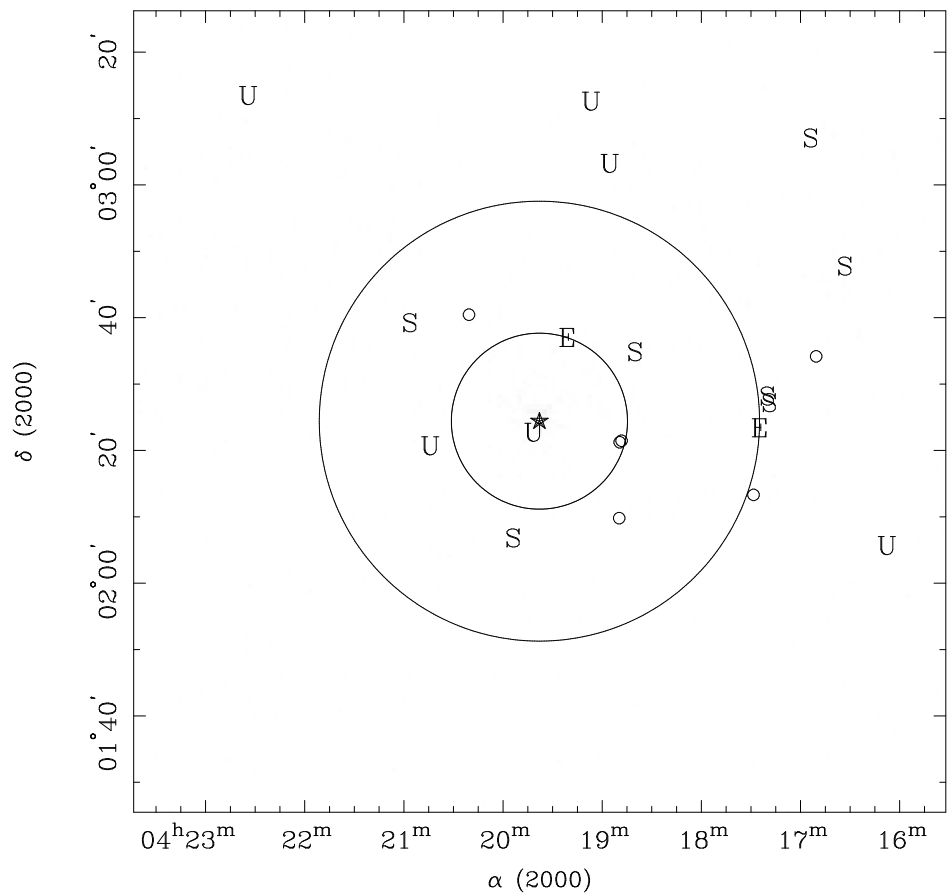}}
  \caption{The projected galaxy distribution in the NGC~1550 group. The central star
represents NGC~1550. Early-type member galaxies are represented by `E' (including
Sa galaxies), while the late-type member galaxies are marked by `S'. Unclassified
member galaxies are marked by `U'. The small circles represent galaxies brighter
than B $\sim$ 16 mag within the virial radius with no velocity information.
The inner large circle represents 0.2 \rv (or $\sim$ 200 kpc), while the outer
large circle represents 0.5 \rv. All 16 known member galaxies within \rv are marked.
    \label{fig:img:smo}}
\end{figure*}
\clearpage

\begin{figure*}
  \centerline{\includegraphics[height=1.2\linewidth,angle=270]{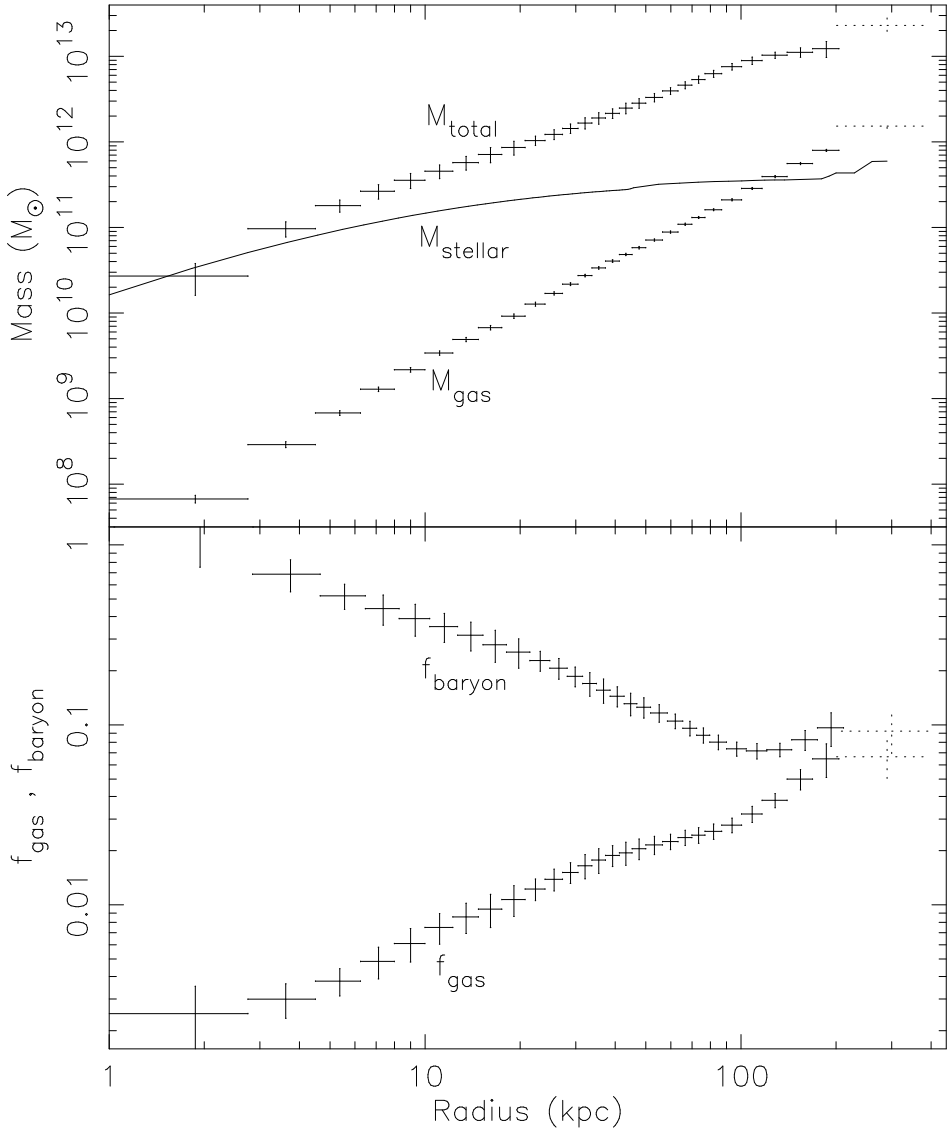}}
  \caption{{\bf Upper}: the total gravitational mass, gas mass and stellar mass
profiles. The jumps of stellar mass profile at large radii are caused by the
inclusion of lights from other luminous member galaxies other than NGC~1550.
The estimated mass values at $\sim$ 300 kpc are shown in dotted
lines. They have much larger statistical and systematic uncertainties
than derived values in smaller radii; {\bf Lower}: the gas fraction and
the baryon fraction profiles. The uncertainties of the stellar mass are not
included in the uncertainties of the baryon fraction. We slightly shift the
baryon fraction profile along the X position for clarity. The gas fraction
and baryon fraction reach $\sim$ 0.07 and 0.1 respectively at 190 kpc
and show signs of flattening beyond this radius.
    \label{fig:img:smo}}
\end{figure*}
\clearpage

\begin{figure*}
  \centerline{\includegraphics[height=0.9\linewidth,angle=270]{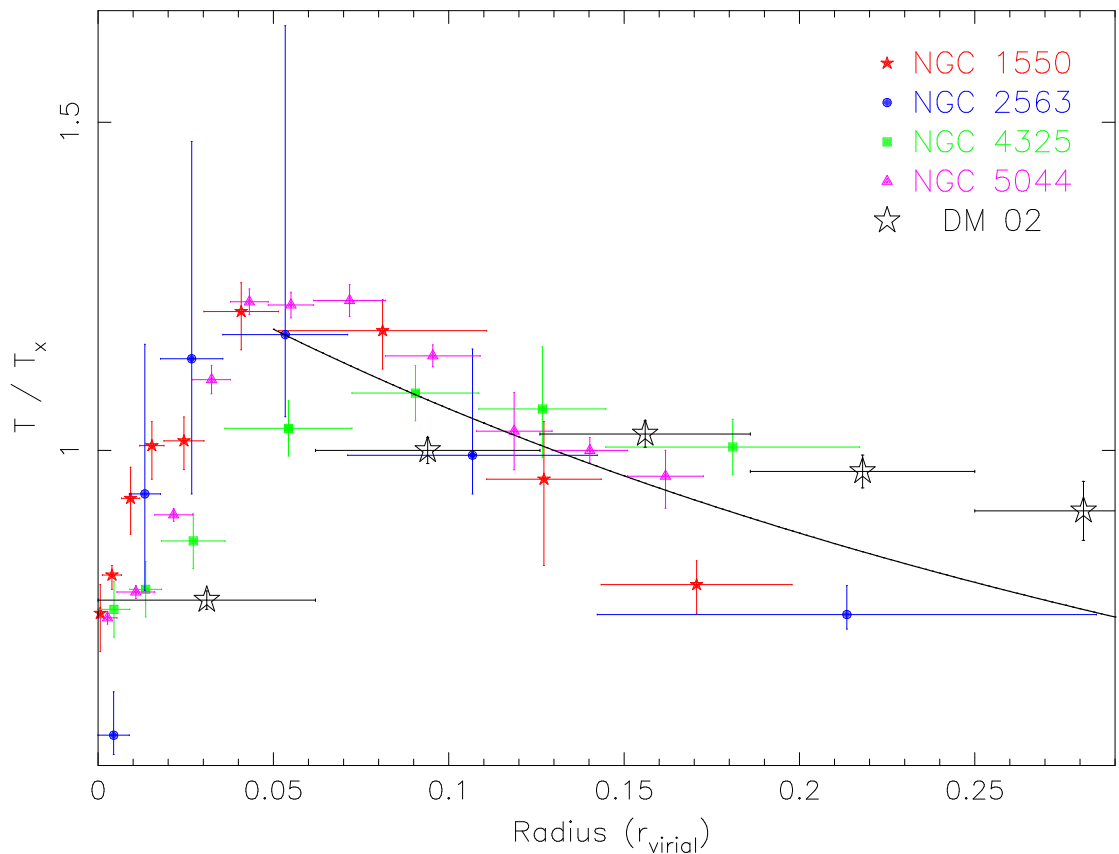}}
  \caption{Normalized deprojected temperature profiles of the four 1 keV galaxy groups
compared with temperature profiles of hot ($>$ 3.5 keV) clusters. The large
stars in black represent the \sax\ projected temperature profile derived by DM02.
The temperature profile of M98 is cooling-flow corrected and cannot be
directly compared to those with single temperature fits. The solid
line represents the ``Universal temperature profile'' (projected) derived in
adiabatic simulations by Loken et al. (2003). The mean temperatures obtained
from MEKAL fits to the global spectra of the four 1 keV groups (NGC~1550,
NGC 2563, NGC 4325 and NGC 5044) are 1.37, 1.36, 0.95 and 1.02 keV
respectively (this work; M03; Buote et al. 2003a).
    \label{fig:img:smo}}
\end{figure*}
\clearpage

\begin{figure*}
  \centerline{\includegraphics[height=0.9\linewidth,angle=270]{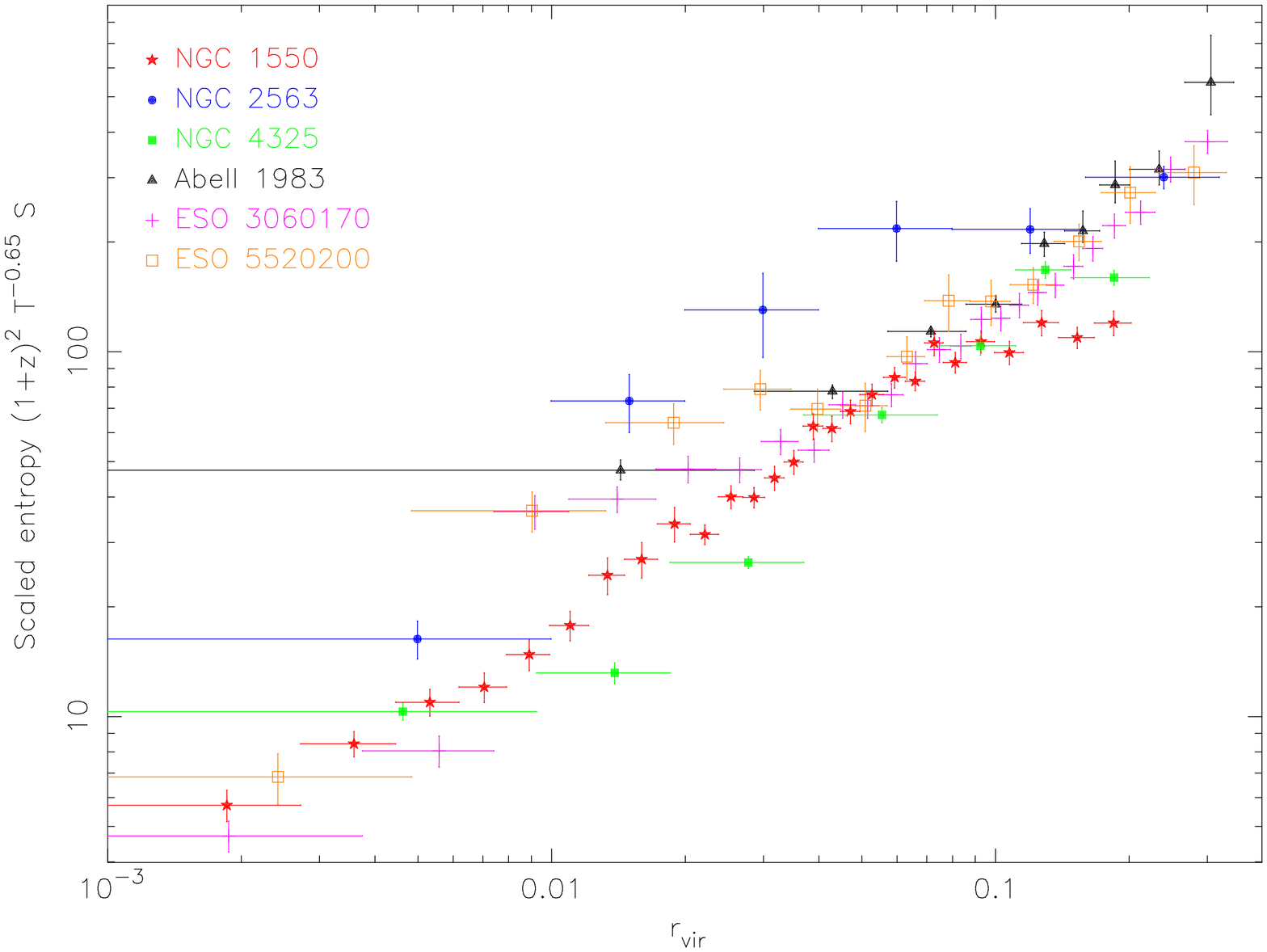}}
  \caption{Scaled entropy profiles of 6 galaxy groups with average temperatures
from 0.95 to 2.7 keV. The scaled entropy profiles of 2 - 3 keV groups (Abell 1983,
ESO 3060170 and ESO 5520200) align with each other well, while the scaled entropy
profiles of 1 keV groups (NGC 1550, NGC 2563 and NGC 4325) show larger scattering.
The scaled entropy profiles of these 1 keV groups flatten between $\sim$ 0.1
and 0.2 \rv, but with different values spanning almost a factor of three.
    \label{fig:img:smo}}
\end{figure*}


\begin{references}
 \reference{} Allen, S. W., Ettori, S. \& Fabian, A. C. 2001, MNRAS, 324, 877
 \reference{} Allen, S. W., Schmidt, R. W., \& Fabian, A. C. 2001, MNRAS, 328, L37
 \reference{} Anders, E., \& Grevesse N. 1989, Geochimica et Cosmochimica Acta, 53, 197
 \reference{} Arnaud, M. et al. 2001, A\&A, 365, L80
 \reference{} Babul, A., Balogh, M. L., Lewis, G. F., \& Poole, G. B. 2002, MNRAS, 330, 329
 \reference{} Beers, T. C., Flynn, K., Gebhardt, K. 1990, AJ, 100, 32
 \reference{} Beuing, J., D\"obereiner, S., B\"ohringer, H., \& Bender, R. 1999, MNRAS, 302, 209
 \reference{} Borgani, S. et al. 2002, MNRAS, 336, 409
 \reference{} Buote, D. A. 2000, ApJ, 539, 172
 \reference{} Buote, D. A., Lewis, A. D., Brighenti, B., Mathews, W. G. 2003a, ApJ, in press (astro-ph/0205362) 
 \reference{} Buote, D. A., Lewis, A. D., Brighenti, B., Mathews, W. G. 2003b, ApJ, in press (astro-ph/0303054)
 \reference{} Cole, S. et al. 2001, MNRAS, 326, 255
 \reference{} Dav$\acute{\rm e}$, R., Katz, N., \& Weinberg, D. H. 2002, ApJ, 579, 23
 \reference{} David, L. P., Forman, W., \& Jones, C. 1991, ApJ, 380, 39
 \reference{} David, L. P., Jones, C., Forman, W., Daines, S. 1994, ApJ, 428, 544
 \reference{} David, L. P., Jones, C., \& Forman, W. 1995, ApJ, 445, 578
 \reference{} Davis, D. S., Mulchaey, J. S., Mushotzky, R. F., Burstein, D. 1996, ApJ, 460, 601
 \reference{} De Grandi, S., \& Molendi, S. 2002, ApJ, 567, 163 (DM02)
 \reference{} Evrard, A. E.; Metzler, C. A.; Navarro, J. F. 1996, ApJ, 469, 494 (EMN96)
 \reference{} Finoguenov, A., David, L. P., Ponman, T. J. 2000, 544, 188
 \reference{} Finoguenov, A., Arnaud, M., David, L. P. 2001, ApJ, 555, 191
 \reference{} Finoguenov, A., Jones, C., B\"ohringer, H., Ponman, T. J. 2002, ApJ, 578, 74
 \reference{} Garcia, A. M. 1993, A\&AS, 100, 47
 \reference{} Gibson, B. K., Loewenstein, M., \& Mushotzky, R. 1997, MNRAS, 290, 623
 \reference{} Giuricin, G., Marinoni, C., Ceriani, L., Pisani, A. 2000, ApJ, 543, 178
 \reference{} Helsdon, S. F., \& Ponman, T. J. 2000, MNRAS, 315, 356 (HP00)
 \reference{} Kauffmann, G. et al. 2003, MNRAS, 341, 33
 \reference{} Kochanek, C. S. et al. 2001, ApJ, 560, 566
 \reference{} Kraft, R. P. et al. 2003, ApJL, submitted (astro-ph/0304362)
 \reference{} Lloyd-Davies, E. J., Ponman, T. J., \& Cannon, D. B. 2000, MNRAS, 315, 689
 \reference{} Loken, C. et al. 2003, ApJ, 579, 571L 
 \reference{} Majerowicz, S., Neumann, D. M., \& Reiprich, T. H. 2002, A\&A, 394, 77
 \reference{} Markevitch, M., Forman, W. R., Sarazin, C. L., \& Vikhlinin, A. 1998, ApJ, 503, 77 (M98)
 \reference{} Markevitch, M. 1998, ApJ, 504, 27
 \reference{} Markevitch, M., Vikhlinin, A. 2001, ApJ, 563, 95
 \reference{} Markevitch, M., et al. 2003, ApJ, 583, 70
 \reference{} Moore, B., Governato, F., Quinn, T., Stadel, J., \& Lake, G. 1998, ApJ, 499, L5
 \reference{} Muanwong, O., Thomas, P. A., Kay, S. T., \& Pearce, F. R. 2002, MNRAS, 336, 527
 \reference{} Mulchaey, J. S., \& Zabludoff, A. I. 1998, ApJ, 496, 73 
 \reference{} Mulchaey, J. S., Davis, D. S., Mushotzky, R., Burstein, D. 2003, ApJS, in press (astro-ph/0302393)
 \reference{} Mushotzky, R., Figueroa-Feliciano, E., Loewenstein, M., Snowden, S. L. 2003, astro-ph/0302267 (M03)
 \reference{} Navarro, J., Frenk, C., \& White, S. 1997, ApJ, 490, 493
 \reference{} Ponman, T. J., Cannon, D. B., \& Navarro, J. F. 1999, Nature, 397, 135
 \reference{} Ponman, T. J., Sanderson, A. J. R., \& Finoguenov, A. 2003, MNRAS, in press (astro-ph/0304048)
 \reference{} Pratt, G. W., Arnaud, M., Aghanim, N. Clusters of galaxies and the high
redshift universe observed in X-rays, Recent results of XMM-Newton and Chandra,
XXXVIth Rencontres de Moriond , XXIst Moriond Astrophysics Meeting, 2001, edited by D. M.
Neumann \& J. T. T. Van, 38
 \reference{} Pratt, G. W., \& Arnaud, M. 2002, A\&A, 394, 375
 \reference{} Pratt, G. W., \& Arnaud, M. 2003, A\&A, in press (astro-ph/0304017)
 \reference{} Sanderson, A. J. R., Ponman, T. J., Finoguenov, A., Lloyd-Davies, E. J., \& Markevitch, M. 2003, MNRAS, in press, (astro-ph/0301049)
 \reference{} Sun, M., Jones, C., Murray, S. S., Allen, S. W., Fabian, A. C., Edge, A. 2003, ApJ, 587, 619
 \reference{} Tozzi, P., \& Norman, C. 2001, ApJ, 546, 63 (TN01)
 \reference{} Vikhlinin, A. et al. 1999, ApJ, 520, L1
 \reference{} Voit, G. M., \& Bryan, G. L. 2001, Nature, 414, 425
 \reference{} Young, P. J. 1976, AJ, 81, 807
 \reference{} Zabludoff, A. I., \& Mulchaey, J. S. 1998, ApJ, 496, 39
\end{references}
\end{document}